\documentclass[10pt, journal, twocolumn]{IEEEtran}
\usepackage{etex}
\usepackage{inputenc}
\usepackage{amsmath}
\usepackage{amsthm}
\usepackage{amsfonts}
\usepackage{amssymb}
\usepackage{graphicx}
\usepackage{graphics}
\usepackage{float}
\usepackage{tikz}
\usetikzlibrary{shapes, snakes, patterns}
\makeatletter
\newcommand*{\rom}[1]{\expandafter\@slowromancap\romannumeral #1@}
\renewcommand{\vec}[1]{\mathbf{#1}}
\makeatother
\usepackage{epstopdf}
\usepackage[export]{adjustbox}
\usepackage[list=true]{subcaption}
\usepackage{color}
\usepackage{algorithm}
\usepackage{algorithmic}

\usepackage[justification=centering]{caption}
\usepackage{enumerate} 
\usepackage{cite}
\usepackage{suffix}
\usepackage{mathtools}
\usepackage{tabularx, booktabs}
\usepackage{multirow}
\usepackage{diagbox}

\definecolor{Gray}{gray}{0.9}

\newcolumntype{C}{>{\centering\arraybackslash}X} 
\usepackage[font=small, labelfont=bf]{caption}
\title{Joint power and resource allocation of D2D communication with low-resolution ADC}
\author{Muralikrishnan Srinivasan$^{1}$, Athira Subhash$^{2}$, Sheetal Kalyani   \\
 \thanks{The authors are with the Department of Electrical Engineering, Indian Institute of Technology Madras,  Chennai, India 600036 (email:\{ee14d206@ee,ee16d027@smail,skalyani@ee\}.iitm.ac.in). }
 \thanks{Muralikrishnan Srinivasan and Athira Subhash are co-first authors.}
 }
\begin{document}

\maketitle
\begin{abstract}
This paper considers the joint power control and resource allocation for a device-to-device (D2D) underlay cellular system with a multi antenna BS employing ADCs with different resolutions.
We propose a four step algorithm that optimizes the ADC resolution profile at the base station (BS) to reduce the energy consumption and perform joint power control and resource allocation of D2D communication users (DUEs) and cellular users (CUEs) to improve the D2D reliability. 
\end{abstract}
\section{Introduction}
Device to device (D2D) communication has gained a lot of interest in recent literature. Here, two mobile users in proximity communicate directly without traversing the eNodeB (base station BS) or the core network \cite{al2014optimal,jameel2018survey,feng2013device}. D2D communication has found numerous applications in vehicular communications \cite{sun2014d2d}, public safety services \cite{gupta2018cluster}, extension of cellular coverage, cellular offloading \cite{andreev2014cellular}, multi-hop relaying \cite{nishiyama2014relay}, etc. In the D2D underlay mode, the regular cellular users (CUES) use dedicated resource blocks, and the D2D communication users (DUEs) share spectrum resources with the CUEs. This further aids larger throughput and lower latency when compared to communication via BS \cite{jameel2018survey}.  However, managing interference in the D2D underlay network is critical. Therefore, resource allocation strategies, which can manage both the D2D links and the cellular links efficiently become imperative.
\par Works such as \cite{wang2017resource,zhang2016radio,wang2018joint,liang2018resource,liang2018graph} have studied the resource allocation and power control problem for D2D underlay cellular networks from multiple perspectives. The authors of \cite{wang2017resource,wang2018joint} consider the problem when each cellular user shares spectrum with a single D2D. Whereas, works such as \cite{liang2018resource, liang2018graph} consider the case of cellular users sharing spectrum with multiple D2D pairs. However, all the above works consider single antenna terminals for both transmission and reception. Several works such as \cite{lin2015interplay,xu2018pilot, shalmashi2015energy, amin2015power} consider D2D underlay cellular systems with multi-antenna BS. One major concern with the multi-antenna system, especially in the massive MIMO system, is the cost and energy consumption of the large number of high-resolution ADCs for the receivers. Motivated by environmental concerns and economic factors, green communication strategies in which the energy consumption at the BS is minimized while satisfying the quality‐of‐service (QoS) requirements of the end users has gained significant traction \cite{bai2015ee}.
\par  Using low-cost and power-efficient low-resolution ADCs, at the cost of signal quality and quantization noise is one way to minimize BS energy consumption \cite{jacobsson2017throughput, fan2015uplink, ding2018outage, xu2019uplink,srinivasan2019analysis}. However, there are no works in the open literature that solves the problem of power control and spectrum allocation of D2D communication in a cellular network taking into account the presence of low-resolution ADCs. Therefore, in this work, we consider a D2D underlay cellular network similar to \cite{liang2018d2d}. In addition, we also consider a multi-antenna BS with low-resolution ADCs with a constraint on the energy consumption at the BS. We solve a joint power-control and spectrum allocation problem using a four step algorithm, in which we optimize the ADC resolution profile at the BS to reduce the energy consumption, perform joint power control of D2D users and cellular users to improve the D2D reliability and allocate spectrum to maximize the ergodic sum rate of the cellular users. 

\section{System Model} \label{system}
We consider the uplink of a D2D-based vehicular communications network in which there are $M$ CUEs (Vehicle-to-interface links) and $K$ DUE pairs (D2D pairs or Vehicle-to-vehicle links). The CUEs are denoted by the set ${\mathcal M} = \{1,\cdots,M\}$ and the DUE pairs by the set ${\mathcal K} = \{1,\cdots,K\}$. We assume that each of the CUEs occupies an orthogonal spectrum and the number of D2D pairs is larger than that of the CUE users, i.e., $K \gg M$. Therefore, spectrum reuse among DUE pairs is necessary. The BS has $N_R$ antennas with low-resolution ADCs. Further, due to the rapidly varying channel with respect to the D2D users, the BS does not have the instantaneous CSI of the DUE pairs. However, the statistical nature and the slow fading parameters of the DUEs are available at the BS. 
\par The channel coefficients from the $m$th CUE to the BS is $\vec g_{m,B} = [g_{m,B}^1,..., g_{m,B}^{N_R}]^T$, where the $i$th element is $g_{m,B}^i = \sqrt{\alpha_{m,B}}h_{m,B}^i, \: i=1,..,N_R.$
Here, $h_{m,B}^i \sim CN(0,1) $, the small-scale fading component, is identical and independent (i.i.d.) across the links and antennas. $\alpha_{m,B}$ denotes the antenna-invariant large-scale fading effects owing to path loss and shadowing. Similarly, the channel between the Tx (transmitter) and the Rx (receiver) of the $k$th DUE pair is $g_{k}$. The channel coefficient from the $k'$th DUE-pair's Tx to the $k$th DUE-pair's Rx is $g_{k',k}$. Also, the interfering channel from the $m$th CUE to the $k$th DUE-pair's Rx is $g_{m,k}$, and the interfering channel from the $k$th DUE-pair's Tx to the BS is
$\vec g_{k,B}= [g_{k,B}^1,..., g_{k,B}^{N_R}]$.
We have,
    $g_k = \sqrt{\alpha_k} h_k, \: g_{k',k} = \sqrt{\alpha_{k',k}}h_{k',k},
    g_{m,k}=\sqrt{\alpha_{m,k}}h_{m,k} \: \text{and} \ 
    g_{k,B}^i= \sqrt{\alpha_{k,B}} h_{k,B}^i$, 
where $\alpha_k$, $\alpha_{k',k}$, $\alpha_{m,k}$ and $\alpha_{k,B}$ are the corresponding slow fading coefficients. Before formulating the power control and spectrum allocation problem, the metrics to be optimized are first discussed.
\subsection{Energy consumption at the BS}
Consider a massive MIMO system, i.e., $N_R$ is large. Let the resolution of the ADC for the $i$th BS antenna be $b_i$ bits. A mixed ADC-structure is considered here, where $b_i$ can take any integer value from $1$ to $B_{max}$ \cite{ding2018outage}. Let $\vec b =[b_1,...,b_{N_R}]$ represent the ADC resolution profile of the antennas at the BS receiver. The BS energy consumption is \cite{ding2018outage}
\begin{equation}
    E_{BS}= c_0\sum_{i=1}^{N_R}2^{b_i} +c_1,
\end{equation}
where $c_0$ and $c_1$ are coefficients independent of the ADC resolution profile $\vec b$. For $n= 1,..., B_{max}$, let $l_n$ denote the number of antennas whose ADC-resolution is $n$ bits and let $\vec L=[ l_1,..., l_{B_{max}}]^T$. Typically, the energy consumption at the BS has to be optimized such that it is below a threshold.

\subsection{Outage probability at DUE}
The signal-to-interference and noise ratio (SINR) at the $k$th DUE Rx is
\begin{align}
    \gamma_k^d= \frac{P_k^d |{g}_{k}|^2}{\sigma^2+\underset{m \in \mathcal{M}}{\sum}\rho_{m,k} P_m^c |{g}_{m,k}|^2 + \sum\limits_{k'\ne k}\rho_{m,k'} P_{k'}^d |g_{k',k}|^2 },
\end{align}
where $\sigma^2$ is the noise power,  $P_m^c$ and $P^d_k$ denote transmit powers of the $m$th CUE and the $k$th DUE Tx, respectively. Here, $\rho_{m,k}=1$ indicates that the $k$th DUE-pair reuses the spectrum of the $m$th CUE and $\rho_{m,k}=0$ otherwise. To guarantee reliability of the D2D links, the outage probability at the DUE has to be below a tolerable threshold $p_0$. In other words, $\text{Pr}\left\{\gamma_{k}^d \le \gamma_{0}^d \right\} \le p_0$, where $\gamma_{0}^d$ is the SINR threshold.

\subsection{Ergodic rate of the CUEs}
To determine the ergodic rate of the CUEs at the BS, we first consider the signal from the $m^{th}$ CUE received at the BS antenna at a given time instant,
\begin{equation}
    \vec{y} = P_m^c\vec{g}_{m,B}x_m+\sum\limits_{k \in \mathcal{K}}\rho_{m,k}P_k^d \vec{g}_{k,B}x_k + \vec{n},
\end{equation}
where $x_m$ and $x_k$ are the symbols transmitted from the $m^{th}$ CUE and $k^{th}$ DUE respectively and $n$ is the additive white Gaussian noise with variance $\sigma^2$. Now, owing to the quantization noise from the low resolution ADC the received signal at the BS is given by \cite{ding2018ee} 
\begin{equation}
    \vec y_q = \vec A_b \vec y + \vec n_q,
\end{equation}
where $\vec A_b =\text{diag}(a_{b_1}, a_{b_2},..., a_{b_{N_R}})$ models the quantization profile and $\vec n_q$ is the additive quantization noise vector that is uncorrelated with $\vec y$. The $i$th diagonal entry of $\vec A_b$, given by $a_{b_i}$ is related to the number of quantization bins by \cite[Table I]{ding2018outage} for $b_i \leq 5$ and by the relation $a_i = 1-\frac{\pi \sqrt{3}}{2}2^{-2b_i}$ for $b_i > 5$.  After maximum ratio combining (MRC), the received signal $r$ is given by
\begin{equation}
    \vec r = \vec g_{m,B}^H \vec y_q.
\end{equation}
Hence, the interference and noise power affecting the $m^{th}$ CUE signal at the BS is given by,
\begin{align}
    \nonumber
      I_m &= \sum\limits_{k \in \mathcal{K}}\rho_{m,k}P_{k}^d|\vec{g}_{m,B}^H \vec A_b\vec{g}_{k,B}|^2+\sigma^2||\vec{g}_{m,B}^H \vec A_b||^2 \\
    &+ \vec{g}_{m,B}^H \vec R_{n_q, n_q}\vec{g}_{m,B},
\end{align}
where $\vec{R_{n_q n_q}}$ is the covariance of $\vec n_q$ and is given by \cite[Eq. 5]{fan2015uplink}. Therefore, the instantaneous signal-to-interference-quantization noise ratio (SIQNR) at the $m$th CUE is
\begin{align}
    \gamma_m^c= \frac{P_m^c|\vec{g}_{m,B}^H \vec A_b\vec{g}_{m,B}|^2}{I_m}
\end{align}
It is intractable to determine the exact ergodic rate $R_m^c=\mathbb{E}  \left\lbrace  \text{log}_2(1+\gamma_m^c) \right\rbrace$ due to the presence of the quantization noise term \cite{fan2015uplink, srinivasan2019analysis, ding2018outage}. Hence, using the popular approximation $\mathbb{E}\left(\text{log}_2\left(1+\frac{X}{Y}\right)\right) \approx \text{log}_2\left(1+\frac{\mathbb E(X)}{\mathbb E(Y)}\right)$, the ergodic rate is \cite{ding2018ee}
\begin{align}
    R_m^c= \text{log}_2\left(1+ \frac{ P_m^c \alpha_{m,B}^2 (\psi_1^2+\psi_2)}{\nu \psi_1-2 P_m^c \alpha_{m,B}^2 \psi_2} \right)\label{rate_CUE},
\end{align}
where $\psi_1= \sum_{i=1}^{N_R}a_{b_i}$, $\psi_2= \sum_{i=1}^{N_R}a_{b_i}^2$ and $\nu=\sigma^2\alpha_{m,B}+  \sum_{k \in \mathcal{K}}P_k^d \rho_{m,k} \alpha_{m,b} \alpha_{k,B}  + 2 P_m^c \alpha_{m,B}^2$. Ideally, the ergodic rate of the CUEs at the BS have to be maximized.
\section{Spectrum allocation problem and solution}
The spectrum and power allocation problem is formulated in this work as:
\begin{align}
  &\underset{\{\rho_{m,k}\}, \{P_{m}^c\},\{P_{k}^d\}, \vec L}{\max} \sum\limits_{m} R_m^c, \label{eqOpt}\\
 \text{s.t.} ~~~~ & \text{Pr}\left\{\gamma_{k}^d \le \gamma_{0}^d \right\} \le p_0, \forall k, \tag{\ref{eqOpt}a} \\
& \sum\limits_{m}\rho_{m,k} = 1, \: \rho_{m,k} \in \{0,1\}, \forall m,k,\tag{\ref{eqOpt}b} \\
&  0 \le P_{m}^c \le P_{\text{max}}^c, \forall m, \:  0 \le P_{k}^d \le P_{\text{max}}^d, \forall k, \tag{\ref{eqOpt}c} \\
& E_{BS} \leq J, \quad  \sum_{n=1}^{B_{max}}l_n=N_R, \tag{\ref{eqOpt}d}
\end{align}
where $\gamma_0^d$ in (\ref{eqOpt}a) is the minimum SINR guarantee at the DUE-Rxs  and $p_0$ in (\ref{eqOpt}a) is the maximum tolerable outage probability. Constraint (\ref{eqOpt}a) represents the minimum reliability requirement for the $K$ DUE links. Constraint (\ref{eqOpt}b) ensures that each DUE pair uses spectrum of only one CUE link. Constraint (\ref{eqOpt}c) ensures that the transmit powers of CUE and DUE links have maximum limits $P_{\text{max}}^c$ and $P_{\text{max}}^d$ respectively.
Constraint (\ref{eqOpt}d) ensures that the energy consumption at the BS is below a threshold $J$. 
\par It is the presence of constraints (\ref{eqOpt}d) and use of multi-antenna receiver at the BS in the presence of quantization error that makes our problem different from those of existing works like \cite{liang2018d2d, wang2018joint}. Although employing a higher resolution ADCs reduces the quantization error and improves the CUE rate, a higher power dissipation of the ADC is also incurred and constraint (\ref{eqOpt}d)  cannot be met. Therefore, determining the right trade-off between the cellular rate and energy consumption at BS is essential. Hence, it is imperative that the ADC resolutions of a multi-antenna receiver are optimized with the goal of minimizing the energy consumption at the BS and maximizing the CUE rate.
\par Obtaining an optimal solution to the spectrum and power allocation problem is complicated due to presence of integer constraints (\ref{eqOpt}d) and the objective function that involves the quantization term. Here, a sub-optimal solution that can be centrally implemented at the BS in four steps is proposed. Hence the name 4-step algorithm (4SA) is used henceforth. First, the resolution profile $\vec L$ that maximizes the ergodic sum rate and satisfies the constraints in (\ref{eqOpt}d) simultaneously is determined. Second, since the DUEs reuse spectrum, the DUEs is clustered such that the intra-cluster interference is minimized. Third, for a given ADC resolution profile and DUE clustering, the power allocation problem for every pair of CUE and DUE cluster is solved. Finally, resource matching is performed such that the CUE ergodic sum rate is maximized.
\subsection*{Step 1: Determining resolution profile}
Since the ergodic sum rate of the CUEs, $R_m^c$, and the energy consumption at the BS, $E_{BS}$ are dependent on the resolution profile $\vec L$, we have to design a resolution profile $\vec L$ (optimize the number of ADCs for each resolution value) such that $R_m^c$ is maximized and $E_{BS}$ is minimized. This is an integer programming problem with a non-linear objective function, a linear constraint and an integer constraint (\ref{eqOpt}d). The complexity of an exhaustive search is $O(N_R^{ B_{max}})$, which is costly for large $N_R$. Note that an incremental search algorithm with polynomial complexity is proposed in \cite{ding2018outage} to minimize BS energy consumption for an outage probability constraint of the users at the BS. Inspired from \cite{ding2018outage}, we propose a decremental search algorithm to determine the optimum resolution profile.
\par The BS have the lowest energy consumption and and the CUEs have the lowest ergodic rate when all the ADCs are $1$-bit ADCs. Also, the  $E_{BS}$ and $R_m^c$ monotonically increases with respect to each ADC resolution. To begin with, the vector $\vec L$ is chosen to match the case with the highest $R_m^c$ and $E_{BS}$. This corresponds to $\vec L= [\vec 0_{B_{max}-1}, N_R]^T$. In other words, we have all $B_{max}$-bit ADCs. Let, $x$, the value of the ADC-resolution level in the current searching step be initialized to $B_{max}$. Let the vector at the current searching step be $\vec L_c$ and its $E_{BS}$ be $E_{BS}(c)$. In each searching step, the number of ADCs with resolution level $x$ is decreased by $1$ and $(x-1)$ possible vectors for $\vec L_c$ are tested by increasing one of the ADCs with lower resolutions by 1. The tested vectors at each step is $S_n$
\begin{equation}
   S_n = \{\vec L_c -\vec e_x +\vec e_i , \: \forall i = x-1,...,1\}
\end{equation}
where $\vec e_i$ is the vector in which the $i$th element is $1$ and the other $B_{max}-1$ elements are $0$. Note that $S_n$ is constructed such that as $i$ decreases, $R_m^c$ and $E_{BS}$ decreases. The searching step stops when a vector in $S_n$ that satisfies the $E_{BS}$ constraint is found the first time. If not, the next searching step starts with the same ADC-resolution $x$, after updating $\vec L_c$ with $l_x=l_x-1$ and $l_{x-1}=l_{x-1}+1$. If, $l_x$, the number of antennas with the ADC-resolution $x$ becomes 0, the ADC-resolution $x$ is decreased by $1$ and the procedure repeats till $\vec L$ is determined.
\begin{table}[!t]
\caption{Decremental search algorithm for determining resolution profile $\vec L$ \cite{ding2018outage}} \label{alg:resolution} \centering
\begin{algorithm}[H]
\caption{Algorithm for determining $\vec L$} \label{algm:maxCut}
\begin{algorithmic}[1]
\normalsize{
\STATE \textbf{Initialization:} $x \leftarrow B_{max}$; $\vec L_c \leftarrow [\vec 0_{B_{max}-1}, N_R]^T$
\WHILE  {$x > 1$}
\STATE $l_{c,x}=l_{c,x}-1$; 
\STATE $flag=0$;
\FOR{$k=x-1:-1:1 $}
\STATE $\vec L= \vec L_c$; $l_k=l_k +1$
\IF{$E_{BS}(\vec L) < J$}
\STATE $flag=1$; $\vec L_c =\vec L$; break;
\ENDIF
\STATE $l_k=l_k -1$
\ENDFOR
\IF{ $flag=0$}
\STATE $l_{x-1}=l_{x-1}+1$; $\vec L_c = \vec L$
\ENDIF
\WHILE {$l_x=0$ and $x >1$}
\STATE $x=x-1$
\ENDWHILE
\ENDWHILE
\STATE Return $\vec L_c$.
}
\end{algorithmic}
\end{algorithm}
\vspace{-4mm}
\end{table}
\subsection*{Step 2: Clustering}\label{sec:clustering}
Next, to divide the DUE links into clusters, we use an algorithm that forms a cluster of DUEs based on their slow fading parameters. In this algorithm, each DUE link is modeled as a vertex of a graph. If any two links are mutually interfering, i.e, they belong to the same cluster, the corresponding vertices are joined by an edge, such that the edge weight is the slow fading channel coefficient. Mathematically, the edge weight  $w_{k',k}=\alpha_{k',k}$. Now we have to partition the $K$ vertices into $N$ sets (clusters), $C_1, \cdots, C_N$, where $N\ll K$, such that the the intra-cluster interference across all clusters, i.e., $\sum\limits_n\left(\sum\limits_{k',k\in C_n}w_{k',k}\right)$ is minimized. This is equivalent to the MAX $N$-CUT problem in graph theory \cite{liang2018graph}. A heuristic algorithm for such a clusering is presented in Table \ref{alg:maxCut}.
\begin{table}[!t]
\caption{Heuristic Algorithm for DUE clustering \cite{liang2018graph}} \label{alg:maxCut} \centering
\begin{algorithm}[H]
\caption{Heuristic Algorithm for DUE clustering} \label{algm:maxCut}
\begin{algorithmic}[1]
\normalsize{
\STATE Arbitrarily assign one DUE link to each of the $N$ clusters.
\FOR{$k \in \mathcal{K}$ \textbf{and} not already in any cluster}
\FOR{$n=1:N$}
\STATE Compute the increased intra-cluster interference using $\sum\limits_{k'\in C_n} (w_{k,k'} + w_{k',k})$.
\ENDFOR
\STATE Assign the $k$th DUE link to the $n^*$th cluster with $n^* = \text{arg} \min \sum\limits_{k'\in C_n} (w_{k,k'} + w_{k',k})$.
\ENDFOR
\STATE Return the DUE clustering result.
}
\end{algorithmic}
\end{algorithm}
\vspace{-8mm}
\end{table}

\subsection*{Step 3: Power Allocation}\label{sec:power}
Once, the DUEs are clustered, for every pair of CUE and a DUE cluster, the CUE ergodic rate is maximized such that the reliability constraints of all the DUE links in that cluster are satisfied. The SIR at the $k$th DUE in the $n$th cluster is
$\gamma_{k,n}^d = \frac{P_{k}^d |g_{k}|^2}{P_{m}^c |g_{m,k}|^2 + \sum\limits_{k'\ne k \in C_n}P_{k'}^d |g_{k',k}|^2 }$.
The power control and rate maximization problem for the $(m,n)$th CUE DUE-cluster pair, where $m$ denotes the CUE index and $n$ denotes the DUE cluster index, is therefore formulated as
\begin{align}
 \max\limits_{P_{m}^c, P_{k}^d} & \text{log}_2\left(1+ \frac{ P_m^c \alpha_{m,B}^2 (\psi_1^2+\psi_2)}{\nu \psi_1-2 P_m^c \alpha_{m,B}^2 \psi_2} \right),  \label{eq1on1} \\
 \text{s.t.} ~~~ &  \text{Pr} \left\lbrace  \gamma_{k,n}^d \le \gamma_{0}^d \right\rbrace  \le p_0, \forall k \in C_n \tag{\ref{eq1on1}a}  \\
& 0 \le P_{m}^c \le P^c_{\text{max}}, \quad  0 \le P_{k}^d \le P^d_{\text{max}}, \forall k \in C_n \tag{\ref{eq1on1}b},
\end{align}
where $\nu=\sigma^2\alpha_{m,B}+ P_k^d \sum_{k \in C_n} \rho_{m,k} \alpha_{m,b} \alpha_{k,B}  + 2 P_m^c \alpha_{m,B}^2$.
For Rayleigh distribution with unit mean power, i.e., $E[|g_k|^2]= 1$, the left hand side of (\ref{eq1on1}a) can be upper bounded by \cite{papndriopoulos2006outage}
\begin{align}
 \text{Pr} \left\lbrace \gamma_{k,n}^d\le \gamma_{0}^d \right\rbrace \leq F_{|g_k|^2}\left(\gamma_0^d\frac{P_m^c \alpha_{m,k}+ \sum\limits_{k'\ne k \in C_n}P_{k'}^d \alpha_{k',k}}{P_k^d \alpha_k} \right).
\end{align}
With this upper bound, the reliability constraints (\ref{eq1on1}a) are transformed to
\begin{align}
 \frac{P_{k}^d\alpha_{k}}{P_{m}^c |g_{m,k}|^2 + \sum\limits_{k'\ne k \in C_n}P_{k'}^d |g_{k',k}|^2} \ge \frac{\gamma_0^d}{F_{|g_k|^2}^{-1}(p_0)}, \forall k \in C_n \label{eq1on1Upper} 
\end{align}
Note that (\ref{eq1on1Upper}) ia a linear inequality constraint in $P_{m}^c$ and $P_{k}^d$, $\forall k \in C_n$. Since the objective (\ref{eq1on1}) is monotonically decreasing with $P_{k}^d$, $\forall k \in C_n$, it can be trivially shown that the objective is maximized only if the outage constraints are satisfied with equality. Therefore, the optimal solution to \eqref{eq1on1} is given by \cite{Sun2017}
\begin{align}\label{eq:optPc}
P_{m}^{c^*} = \min \left\{P_{\text{max}}^c, \left\{ \frac{P^d_{\text{max}} -\bar{\gamma_0}\sigma^2 \pmb{\phi}_i^H\pmb{1}}{\bar{\gamma_0}\pmb{\phi}_i^H\pmb{\alpha}_m} \right\}_{i=1}^{N_{c_n}} \right\},
\end{align}
and
\begin{align}\label{eq:optPd}
\mathbf{P}_{n}^{d^*} = \mathbf{\Phi}^{-1} \bar{\gamma}_0 \left( P_{m}^{c^*} \pmb{\alpha}_m +\sigma^2 \right),
\end{align}
where $\mathbf{P}_{n}^d\in\mathbb{R}^{N_{c_n}\times 1}$ denotes the transmit powers of all $N_{c_n}$ DUE links in the $n$th cluster, $\bar{\gamma}_0 = \frac{\gamma_0^d}{F_{|g_k|^2}^{-1}(p_0)} =\frac{\gamma_0^d}{-ln(1-p_0)}$, $\pmb{\alpha}_m = (\alpha_{m,1}, \cdots, \alpha_{m,N_{c_n}})^T \in \mathbb{C}^{N_{c_n}\times 1}$ and $\mathbf{1}$ is an all-one vector. Also, the $ij$th element of $\mathbf{\Phi}\in\mathbb{C}^{N_{c_n}\times N_{c_n}}$,
is given by $\alpha_i$ if $i=j$ and $-\bar{\gamma}_0\alpha_{j,i}$ if $i \neq j$. $\pmb{\phi}_i^H$ is the $i$th row of $\pmb{\Phi}^{-1}$.
\subsection*{Step 4: Spectrum Allocation}\label{sec:spectrum}
With the optimal power allocations
$(P_{m}^{c^*}, \vec P_{n}^{d^*})$, we aim to maximize the cellular ergodic rates by searching over all possible channel allocation schemes. We first determine the ergodic rate of $m$th CUE given by $R^m_c$ when it shares spectrum with $n$th DUE cluster for a power allocation $(P_{m}^{c^*}, \vec P_{n}^{d^*})$ using (\ref{rate_CUE}). 
The spectrum allocation problem therefore becomes
\begin{align}
\nonumber
   & \underset{\rho_{m,n}\in \{0,1\}}{\text{max}} \sum_{m \in \mathcal{M}}R_m^c,\\
   &\sum_{\forall m \in \mathcal{M}} \rho_{m,n} \leq 1,\: \forall n \in \mathcal{N}, \: \sum_{\forall n \in \mathcal{N}} \rho_{m,n} \leq 1, \: \forall m \in \mathcal{M}.
\end{align}
This is a maximum weight bipartite matching problem and can be solved by the Hungarian method \cite{liang2017d2d, wang2017joint, wang2018joint}.
\section{Numerical Results} \label{results}
We model a multi-lane freeway that passes through a single cell where the BS is located at its center.  Using a spatial Poisson process, the vehicles are dropped on the roads and the vehicle density is determined by the vehicle speed as in \cite{liang2018graph}. The $M$ CUE and $K$ DUE links are randomly chosen among the vehicles. The simulation parameters and the channel models for V2I/CUE and V2V/D2D links  are same as listed in \cite[Table IX]{liang2018graph} \cite[Table X]{liang2018graph} respectively. The number of V2V/D2D clusters, $N$, is set equal to the number of V2I/CUE links, $M$.
\begin{figure}[!h]
\centering
\includegraphics[scale=0.45]{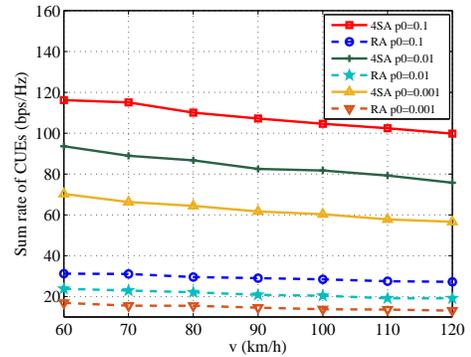}
\caption{Sum of ergodic rate of CUEs vs vehicle speed for $M=10$, $K=30$, $N_R=32$, $J=0.5$}
\label{fig1}
\end{figure}
\par Fig. \ref{fig1} shows the ergodic sum rate of all CUEs with an increasing vehicle speed. In this figure, we compare our 4-step algorithm (4SA) with an ad-hoc random allocation (RA). In RA, the worst case resolution profile is chosen (all the antennas have $1$ bit ADCs) to obtain the least energy consumption and the CUEs are randomly mapped to a DUE cluster. We can observe that our 4SA outperforms RA for all vehicle speed and reliability constraints $p_0$. Note that, higher vehicle speed induces sparser traffic. This, in turn increases the inter-vehicle distance and therefore, the DUE links becomes less tolerable to outage. To compensate for the reduced reliability, the DUEs links have to be allocated more power, which increases the DUE interference at the CUEs and in turn reduces the CUE rates. The sum rates of CUEs achieved becomes larger if the outage probability $p_0$ at the DUEs is increased. This is due to the fact that higher acceptable outage of DUEs increases their tolerance to interference from the CUEs, thereby promoting CUEs to increase their transmit powers.
\begin{table}
\caption{Table: CUE rates at $v=80 km/hr$ and $p_0=0.01$ for various $N_R$, $J$} \label{tbl:threshold}
\begin{center}
\begin{tabular}{ |c||c|c|c| }
\hline
$J$ & $N_R=16$ & $N_R=32$ & $N_R=64$\\
\hline
4  &80.2 & 88.2 & 97.0 \\ 
2 & 80.2& 88.2& 97.0 \\ 
1 & 80.2& 88.2 & 95.8 \\ 
1/2& 80.2 & 86.5& 93.4 \\ 
1/4& 78.7 & 84.3& 89.4 \\ 
1/8&  74.8& 80.2& 83.8 \\ 
1/16&  68.5 & 71.2 & 71.4 \\
1/32& 54.9  & 52.5 & 50.3\\
1/64& 33.1  & 41.6 & 50.3\\
\hline
\end{tabular}
\end{center}
\vspace{-4mm}
\end{table}
\par In TABLE \ref{tbl:threshold}, we give the CUE rates for various values of $J$ and $N_R$. Due to diversity gain, the CUE rates increase on increasing the number of antennas at the BS for large $J$. In this regime, the BS can use high resolution ($B_{max}$ bits) ADCs for all antennas to maximize the CUE rate. With decrease in $J$, the BS is encouraged to use more low-resolution ADCs, thereby decreasing CUE rates. Also, for some low values of $J$, say $J=\frac{1}{32}$, $N_R=16$ outperforms $N_R=32$ or $N_R=64$. This is because, for $N_R=32$ or $64$ the energy consumption can be maintained below $J$ only by equipping all antennas with $1$ bit ADCs. Whereas, for $N_R=16$, one can afford a few ADCs which are greater than $1$ bit and still be below the threshold $J$. If $J$ is further decreased, say for $J=\frac{1}{64}$, all antennas for the case of $N_R=16$ have to be equipped with $1$ bit ADCs. In this regime, due to diversity gain a better CUE rate is obtained for $N_R=64$. 

\section{Conclusion} \label{conclusion}
This paper outlines a 4-step algorithm for power allocation and resource allocation for a D2D underlay cellular system in a multi-antenna system with low resolution ADCs. The optimal ADC resolution profile was first determined such that energy consumption at the BS was minimized. Next the DUEs were clustered such that the intra-cluster interference was minimized. The transmit power of the CUEs and DUEs were then determined such that the maximize the ergodic sum rate of CUEs such that reliability constraints of the D2Ds are met. Finally, resource matching was performed using Hungarian algorithm. Numerical results were shown to understand the trade-off between number of low resolution ADCs, number of BS antennas and energy constraints.

\bibliographystyle{IEEEtran}
\bibliography{library}
\end{document}